\documentclass[10pt, conference, letterpaper]{IEEEtran}
\usepackage{amsfonts}
\usepackage{mathrsfs}
\usepackage{amssymb}
\usepackage{wrapfig}
\usepackage{amsmath}
\usepackage{accents}
\usepackage{dsfont}
\usepackage{amsbsy}
\usepackage{setspace}
\usepackage{graphicx,cite}
\usepackage{subfig}
\usepackage{url}
\usepackage{color}
\usepackage{epstopdf}
\usepackage{amssymb}
\usepackage{framed}
\usepackage{amsthm}

\PassOptionsToPackage{bookmarks={false}}{hyperref}
\usepackage[lined, boxed, linesnumbered, commentsnumbered, ruled]{algorithm2e}
\usepackage[top=0.7in, bottom=1.04in, left=0.632in, right=0.632in, textwidth=7.25in,textheight=9.25in]{geometry}

\IEEEoverridecommandlockouts
\title{Checks and Balances: A Low-complexity High-gain Uplink Power Controller for CoMP}

\author{Fangzhou Chen$^\dag{}^\S$,  Yin Sun$^\dag{}^\S$, Yiping Qin$^\ddag$, and C. Emre Koksal$^\dag$\\
$^\dag$Dept. of ECE, The Ohio State University, Columbus, OH\\
$^\ddag$Huawei Technologies Co., Shanghai, China \\
$^\S$Co-primary authors
\thanks{This work was supported in part by Huawei, Inc. under Agreement YB 2013110091, National Science Foundation under Grants CNS-1054738 and CNS-1514260.}
}

\begin{document}

\renewcommand{\baselinestretch}{1}

\maketitle

\begin{abstract}
Coordinated Multipoint (CoMP) promised substantial throughput gain for next-generation cellular systems. However, realizing this gain is costly in terms of pilots and backhaul bandwidth, and may require substantial modifications in physical-layer hardware. Targeting efficient throughput gain, we develop a novel coordinated power control scheme for uplink cellular networks called Checks and Balances (C\&B), which \emph{checks} the received signal strength of one user and its generated interference to neighboring base stations, and \emph{balances} the two. C\&B has some highly attractive advantages: C\&B (i) can be implemented easily in software, (ii) does not require to upgrade non-CoMP physical-layer hardware, (iii) allows for fully distributed implementation for each user equipment (UE), and (iv) does not need extra pilots or  backhaul communications.
We evaluate the throughput performance of C\&B on an uplink LTE  system-level simulation platform, which is carefully calibrated with Huawei. Our simulation results show that C\&B achieves much better throughput performance, compared to several widely-used power control schemes. 

\end{abstract}
\begin{IEEEkeywords}
Uplink power control, coordinated multipoint (CoMP), LTE, system-level simulation, throughput improvement
\end{IEEEkeywords}

\section{Introduction}
Next generation of cellular communication, e.g., Long Term Evolution Advanced (LTE-A) and 5G, is expected to significantly improve the average throughput and cell-edge throughput for serving user equipments (UEs).
One important candidate technique for achieving such throughput improvement is Coordinated Multipoint (CoMP), which refers to the cooperation among different Base Stations (BSs).

The promised benefits of CoMP are hard to realize because of many issues in practical systems \cite{Lozano13, Suh11}. In particular, the uplink CoMP techniques (such as distributed interference cancellation/alignment, joint detection, coordinated scheduling) all require nearby BSs to communicate received signals, control messages, and channel state information through backhaul links. In addition, these CoMP techniques are  costly in terms of power and pilot resources, which considerably decreases the  resources allocated for data transmissions. Therefore, the realized throughput performance is greatly degraded.
\begin{table}[h!]
\centering
\caption{Performance  of FPC \cite{Whitehead}, Max Power \cite{Furuskar08}, RLPC \cite{Rao07} and C\&B in Macrocell system-level simulations.}
\resizebox{\columnwidth}{!}{
\begin{tabular}{|| c || c |  c | c | c ||}
 \hline
         & FPC & Max Power & RLPC & C\&B\\ 
 \hline
Average Throughput (Mbits/s) & $8.05$ & $12.01$ & $9.78$ & $12.23$     \\
\hline
$5\%$-Edge Throughput (Mbits/s)& $0.23$  & $0.09$ & $0.22$ & $0.23$   \\                                          
\hline
Power Efficiency (Mbits/J) & $751$ & $6.77$ & $226$ & $387$   \\                                       
\hline
\end{tabular}
}
\label{Table_PerformanceSummary}
\vspace{-3mm}
\end{table}

Aiming to realize the potential benefits of CoMP in practical systems, we propose a novel coordinated power control design for uplink cellular networks. The task of uplink power control is to make the signal received  at the base station sufficiently strong, and in the meanwhile keep the interference generated to nearby base stations not severe. In practice, overly high and  low transmission powers are both harmful. Specifically, increasing the transmission power of one UE can increase its  throughput, but it causes some strong interference to nearby cells, which will degrade the throughput of other UEs. Hence, finding the correct balance between a UE's own performance and its incurred cost to the other UEs is crucial to achieve a satisfying performance.

In practical uplink cellular networks, each BS receiver experiences the interference from hundreds of UEs from neighboring cells. Even if perfect CSI knowledge of the signal and interference channels are available, the optimal power control problem is strongly NP hard \cite{Shah11,Luo08}. To make things worse, the base station in current systems typically estimates the signal channel of its served UEs, but the channel coefficients of interfering UEs are mostly unavailable. These practical limitations make the power control problem even more challenging.

In our research, we develop a low-complexity Coordinated power control scheme that provides significant throughput gains, with minimum cost and modifications. To that end, the following are the contributions of this paper:
\begin{itemize}
\item We develop a novel coordinated power control scheme, named Checks and Balances (C\&B). 
C\&B requires very little information, including the large-scale path loss from one UE to several nearby BSs, coarse power level of co-channel interference, and the throughput vs SINR curve of Adaptive Modulation and Coding (AMC). Based on this information, C\&B \emph{checks} the SNR of one UE and its generated INR to nearby BSs, and \emph{balances} the two. 


\item C\&B has some highly attractive advantages: C\&B (i) can be implemented easily in software, (ii) does not require to upgrade non-CoMP physical-layer hardware, (iii) allows for fully distributed implementation for each UE, and (iv) does not need extra pilots or  backhaul communications.

\item We evaluate the throughput performance of C\&B on a system-level simulation platform for LTE uplink, which is carefully calibrated with Huawei. As shown in Table \ref{Table_PerformanceSummary}, C\&B increases the average throughput by $51.9\%$ over Fractional Power Control (FPC) \cite{Whitehead} and $21.5\%$ over Reverse Link Power Control (RLPC) \cite{Rao07}, and achieves similar cell-edge throughput  with FPC and RLPC. Compared to Max Power Control \cite{Furuskar08}, C\&B increases  the average throughput 
and cell-edge throughput by 1.8\% and  $156\%$, respectively, together with greatly improved  power efficiency.


\end{itemize}

We expect C\&B to achieve even better throughput performance when working with physical-layer CoMP techniques, which will be considered in future work.


\noindent {\bf Related Studies:} Non-coordinated power control is standardized in 3GPP protocols \cite{36.213} and has attracted vast research interests \cite{Gipson96, Furuskar08, Whitehead, Castellanos08, Andrews13, Novlan13, Xiao06, Coupechoux11}. There exist three mainstream schemes: 1) Full Compensation Power Control (FCPC) \cite{Gipson96} allocates transmission power to one UE by making full compensation of its large-scale path loss such that all UEs have the same received signal strength, which results in poor performance in per-cell average throughput and inter-cell interference management. 2) Max Power scheme \cite{Furuskar08} let all UEs transmit at their maximum allowable power. It provides high average per-cell throughput, but performs poorly in power efficiency and throughput of UEs at the cell edge. 3) Fractional Power Control (FPC) \cite{Whitehead} is currently the most widely adopted scheme \cite{Andrews13, Novlan13, Xiao06, Coupechoux11}, which allocates transmission power by making fractional compensation of UEs' large-scale path losses, such that UEs in the interior of one cell have stronger received signal strength than UEs at the cell edge. The key drawback remains in the unsatisfying average throughput per cell.

Two coordinated power control schemes have been proposed \cite{Rao07, Yang09}, which make additional compensation for large-scale path losses from one UE to its neighboring cells. These schemes are essentially variations of FCPC. Hence, they partially inherit the drawbacks of FCPC which significantly limits the throughput gain.

\section{System Model and Problem Description}\label{SystemModel}

We consider an LTE uplink multicellular network. In such network, each UE transmits to its serving cell and meanwhile generates interference to its neighboring cells. Consider UE $u$ transmits signal $x$ at power $P$ on a single subcarrier, its received signal $y$ at the serving cell $c$ can be expressed as:
\begin{eqnarray}
y = \sqrt{P} \cdot h x + \sum_{j} \sqrt{P_j} \cdot h_{j}x_j + n, \label{Eq_ReceivedSignal}
\end{eqnarray}
where $h$ and $h_{j}$ denotes the instantaneous complex channel gain from UE $u$ and $u_j$ (served in neighboring cell $c_j$) to cell $c$, respectively, $n$ denotes the experienced noise. Hence the total inter-cell interference experienced by UE $u$ is $ \sum_{j} \sqrt{P_j} \cdot h_{j}x_j$. Assuming the transmitted signals $x$ and $x_j$ have unit variance, the signal-to-interference-plus-noise ratio (SINR) can thereby be calculated as:
\begin{eqnarray}
sinr =  \frac{P \cdot |h|^2}{ \sum_{j} P_j \cdot |h_{j}|^2 + \sigma_n^2}, \label{Eq_SINR}
\end{eqnarray}
where $\sigma_n^2$ denotes the variance of noise.

The power controller decides the transmission power of UEs across all cells, which heavily affects their SINR. Unlike Non-coordinated power control that only exploits the CSI of UEs to their serving cells, coordinated power control additionally utilizes the CSI of UEs to multiple neighboring cells. The problem we are tackling is to come up with a coordinated power control design with low complexity and high throughput gain over all existing solutions.

Besides the strong impact from power controller, the throughput performance is also influenced by how each cell picks the modulation and coding scheme (MCS) for active UEs. In LTE networks, the existing Turbo-coded modulation techniques are paired with associated bit rate selections to form $29$ different available MCS options \cite{36.213}. For each SINR value, one of these MCS options is chosen by an Adaptive Modulation and Coding (AMC) module. The selected MCS should provide a sufficient high throughput and meanwhile guarantee a low decoding error probability. Usually, the block decoding error rate is required to be less than $10\%$, which will be later compensated by hybrid automatic repeat request (HARQ). 
The stairs in the throughput curve are due to the AMC module. In particular, when the SINR is lower than $-6.5$ dB, no MCS can decode successfully and hence the throughput is zero. When the SINR is higher than $18$ dB, the maximum MCS can decode perfectly, achieving a maximum throughput. Hence, the SINR region for effective AMC selection is  [$-6.5$ dB, $18$ dB].

\section{Checks and Balances: A Power Control Design}\label{Main Design}

In this section, we present a novel power controller design, called Checks and balances (C\&B),  for uplink cellular systems. This power controller requires very little information, including the throughput versus SINR curve of the receiver design, the large-scale path loss from one UE to several nearby BSs, and some coarse distribution information of the co-channel interference in the cellular system. The key idea in C\&B is to cooperatively balance the SNR of a UE and its generated INR to nearby BSs. The complexity of C\&B is very low, and the throughput gain is huge. One can consider C\&B as the simplest implementation of CoMP, which provides significant throughput improvement without incurring huge cost in pilots or backhaul. There is no upgrade of the physical layer design, except for the change of uplink transmission power.

\subsection{Approximations of SINR and Throughput}
C\&B can operate in the open-loop mode, when only large scale CSI is utilized, whereas small scale CSI is unavailable due to the lack of instantaneous channel estimation pilot. The large-scale path loss between UE $u_j$ and cell $c$ is defined as: 
\begin{eqnarray}
\text{PL}_j \triangleq \mathbb{E}\big[|h_j|^2\big]^{-1},  \label{PL}
\end{eqnarray} 
where $\mathbb{E}[\cdot]$ denotes the expectation. When we only consider such large-scale path losses, $u$'s received signal-to-noise ratio (SNR) and interference over thermal noise (IoT) at $c$, and generated interference-to-noise ratio (INR) to $c_j$ can be respectively approximated as:
\begin{eqnarray}
\text{SNR}(P) &=& \frac{\text{PL}^{-1} \cdot P}{N_0}, \label{Eq_SNR} \\
\text{IoT} &=& \frac{N_0 + \sum_{j}{P_j \cdot \text{PL}_{j}^{-1}}}{N_0}, \label{Eq_IoT}\\
\text{INR}_{j}(P) &=& \frac{\text{PL}_{u \to j}^{-1} \cdot P}{N_0}, \label{Eq_INR} 
\end{eqnarray}
where $N_0$ denotes the average noise power and is assumed to be the same and known for all UEs. We also define $\text{PL}_{u \to j}$ as the large-scale path loss from UE $u$ to cell $c_j$. Derived from Eq. (\ref{Eq_SNR}) and (\ref{Eq_IoT}), we can approximate the received SINR of UE $u$ as:
\begin{eqnarray}
\text{SINR} = \frac{\text{SNR}(P)}{\text{IoT}}. \label{Eq_SINR}
\end{eqnarray}

\begin{figure}
 \centering \includegraphics[width=2in]{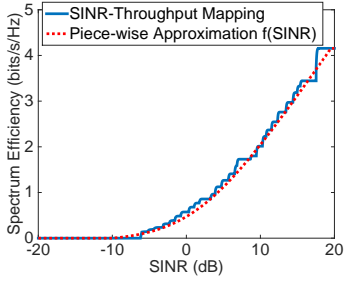} \caption{Throughput Curve Approximation.}
\label{AMC_Approximation} \vspace{-0.2cm}
\end{figure}

As C\&B only acquires large scale CSI and approximated SINR information, there is no need to use an accurate throughput curve to determine the transmission power. Therefore, we introduce a piece-wise function to approximate the foregoing throughput curve:
\begin{eqnarray}\label{U_Basic}
f(\text{SINR}) = \min\big[T_{\max}, a\cdot \log_2(1+b\cdot \text{SINR})\big],
\end{eqnarray}
where $T_{\max} = 4.18$, $a = 0.7035$ and $b = 0.7041$. These parameters are achieved by curve fitting as shown in Fig. \ref{AMC_Approximation}. Note that for different physical layer designs, we can always find such an approximated throughput function $f(\text{SINR})$. Next, we use function $f(\text{SINR})$ to evaluate the throughput of one UE $u$ and the throughput of the UEs interfered by UE $u$.

\subsection{Checking the Influence of SNR and INR} \label{SNR Utility}
\begin{figure}
 \centering \includegraphics[width=2in]{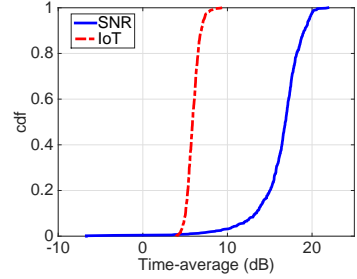} \caption{Time-average SNR and IoT distributions.}
\label{SNR_vs_IoT} \vspace{-0.2cm}
\end{figure}

\begin{figure*}[htp!]
\vspace{-0.2in}
\centerline{
\subfloat[Approximated throughput of UE u.]
{\includegraphics[width=2.1in]{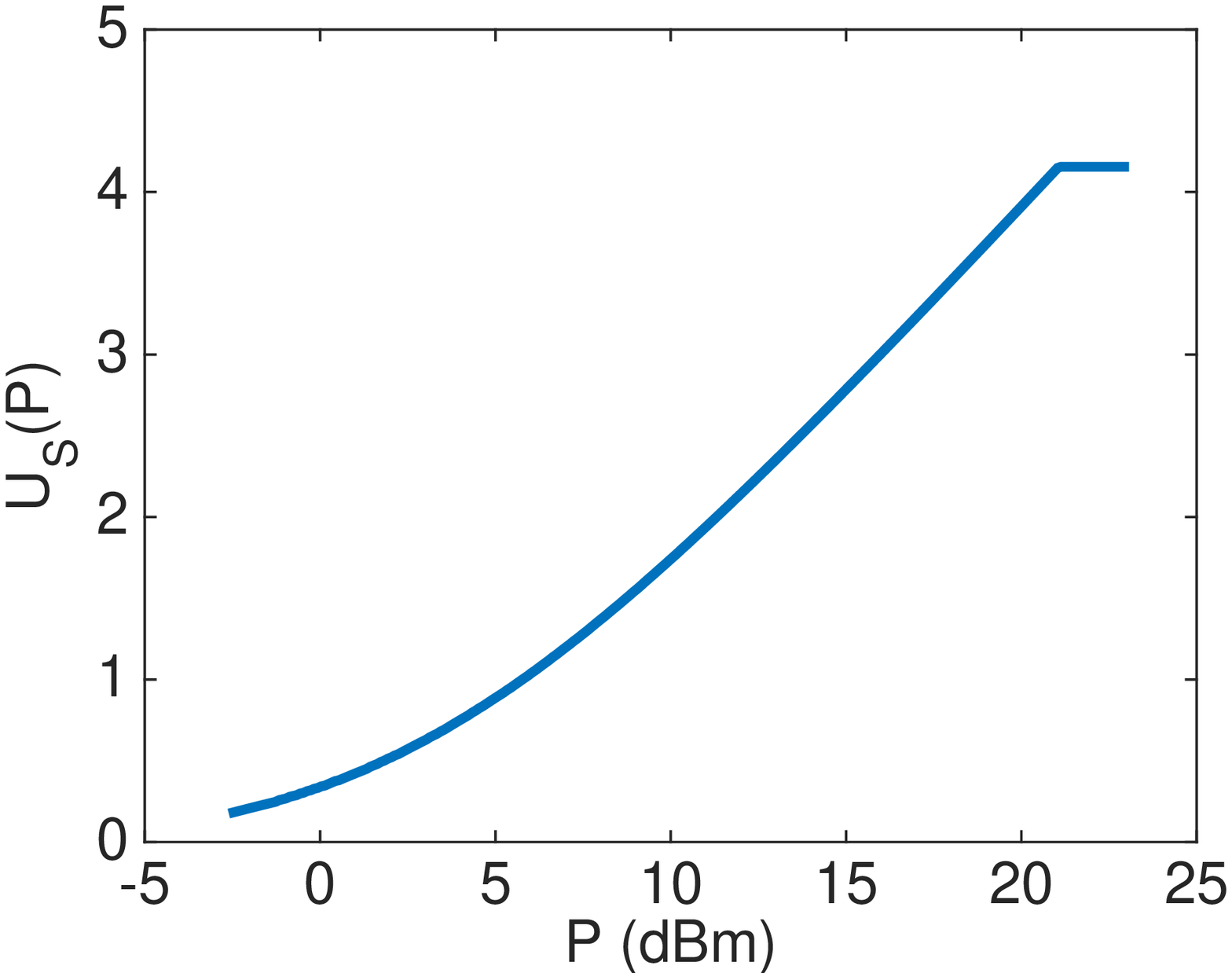}
\label{Sample_Utility_SNR}}
\subfloat[Approximated total throughput of UEs that are interfered by UE $u$.]
{\includegraphics[width=2.1in]{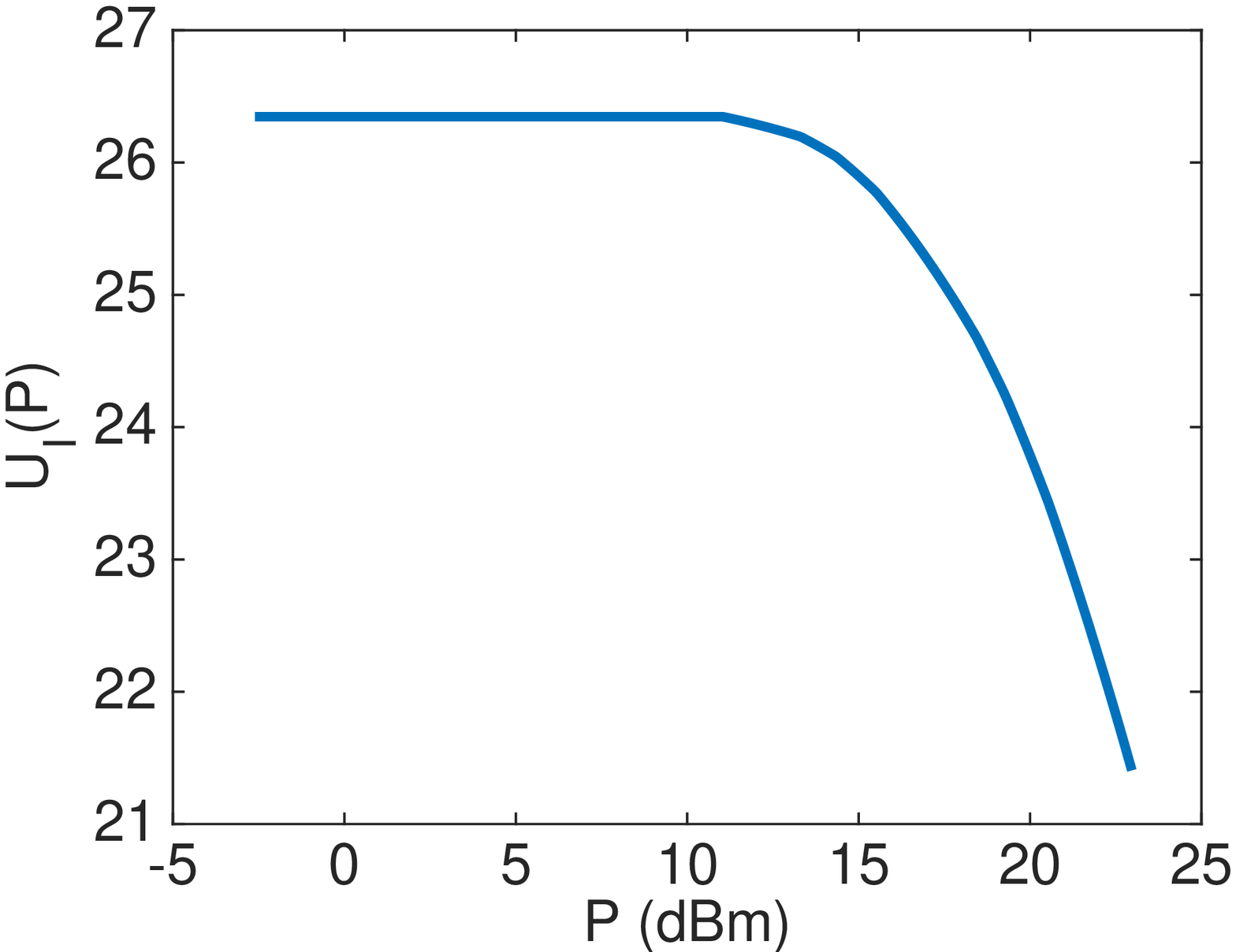}
\label{Sample_Utility_INR}}
\subfloat[Weighted sum throughput approximation.]
{\includegraphics[width=2.1in]{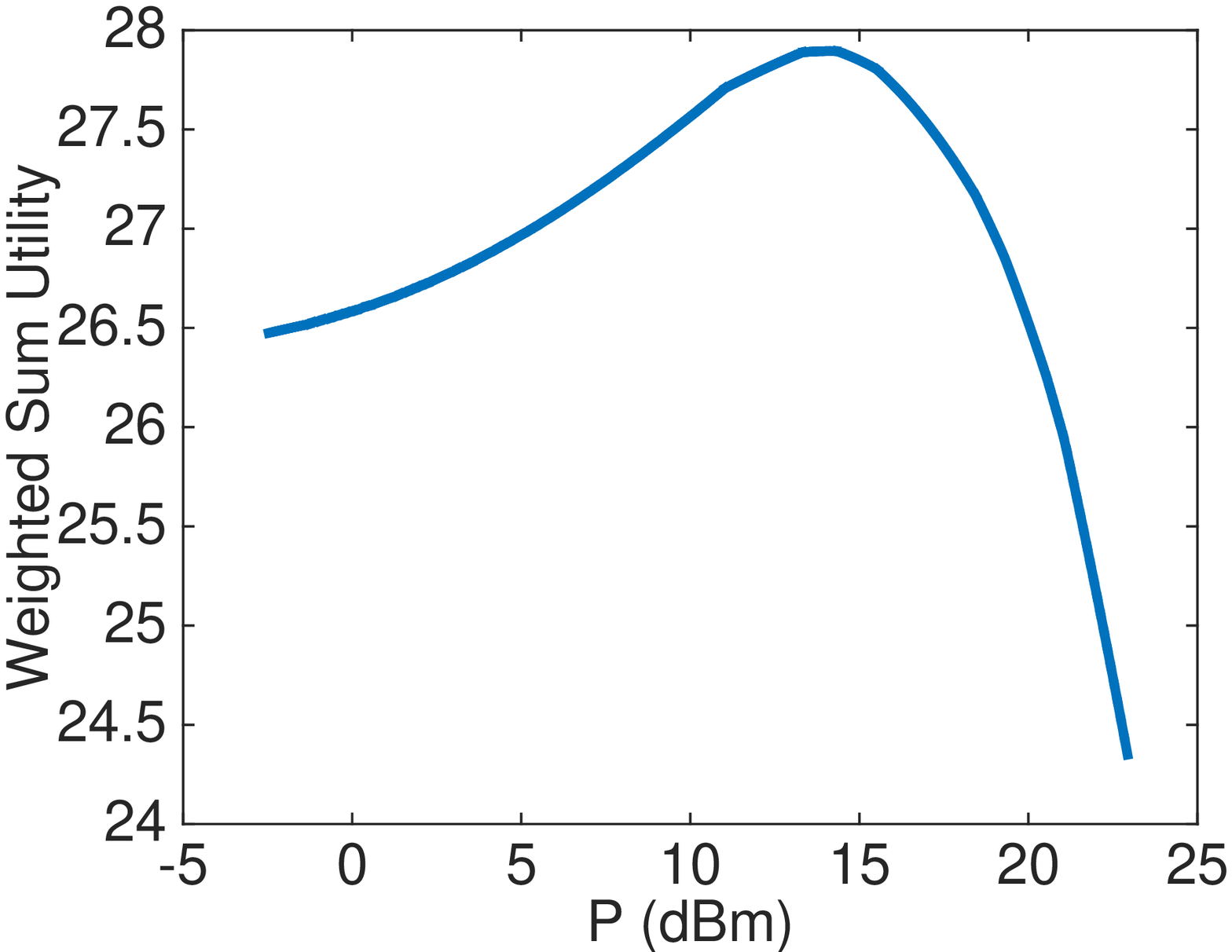}
\label{Sample_Utility_Sum}}
}
\caption{Approximated throughput functions for C\&B power controller design.}
\vspace{-0.2in}
\end{figure*}

Consider an arbitrarily chosen UE $u$, its uplink throughput can be approximated as 
\begin{eqnarray}
R_S(P) = f\left(\frac{\text{SNR}(P)}{\text{IoT}_{S}}\right), \label{U_SNR}
\end{eqnarray}
where $\text{IoT}_S$ denotes a prior estimation of the interference experienced by UE $u$. In fully distributed power control, the instantaneous interference power that UE $u$ will experience during its transmissions is not available at the power controller. Hence, we propose a method to estimate $\text{IoT}_S$. In the simulation environment described in Section \ref{Simulation}, the distribution of the IoT  in the uplink cellular system is illustrated in Fig. \ref{SNR_vs_IoT}. We select  $\text{IoT}_S$ to be the $95^{th}$ percentile of IoT that experienced by an arbitrary UE, which is $\text{IoT}_S=9$ dB. Note that this statistical distribution can be collected at the BS. In practical applications, we recommend to initialize the system with this value and update it according to the measured IoT distribution. The approximated throughput $R_S(P)$ is plotted in Fig. \ref{Sample_Utility_SNR}, which increases with the transmission power $P$.



\noindent


Consider the UEs that are notably interfered by UE $u$, their sum uplink throughput can be approximated as: 
\begin{eqnarray}\label{eq_rate_I}
R_I(P) = \sum_{\text{PL}_{u \to j} < \text{PL}_{th}} f\bigg(\frac{\text{SNR}_I}{\text{IoT}_I + \text{INR}_{j}(P)}\bigg).   \label{U_INR}
\end{eqnarray}
Note that we only consider the non-negligible interference generated by UE $u$. To this end, we set the threshold $\text{PL}_{th}$ in \eqref{eq_rate_I} as the path loss of a interference link such that the interference power generated by UE $u$ with the maximum power $P_{\max}$ is stronger than noise, i.e.,
\begin{eqnarray}
\frac{\text{PL}_{th}^{-1} \cdot P_{\max}}{N_0} > 1, \label{Eq_PLth} 
\end{eqnarray}
where $P_{\max} = 200$ mW ($23$ dBm) \cite{3GPP25.996}. Furthermore, in fully distributed power control, the received SNR and IoT (excluding the interference generated by $u$) of these UEs, denoted by $\text{SNR}_I$ and $\text{IoT}_I$, are unknown to the power controller of UE $u$. We also propose methods to estimate $\text{SNR}_I$ and $\text{IoT}_I$ by using the SINR region [$-6.5$ dB, $18$ dB] in Fig. \ref{AMC_Approximation}  for effective AMC selection. Their values are given by $\text{SNR}_I = 24$ dB and $\text{IoT}_I=5$ dB. Due to space limitations, the details for computing these parameters will be explained in the journal version. The approximated sum throughput $R_I(P)$ is plotted in Fig. \ref{Sample_Utility_INR}, which decreases with the transmission power $P$.

\subsection{Balancing the Two} 
In order to find a proper balance between SNR and INR, we consider the following weighted sum throughput maximization problem:
\begin{eqnarray}
\max_{0 \leq P \leq P_{\max}} R_S(P) + \zeta \cdot R_I(P), \label{U_Maximize}
\end{eqnarray}
where $\zeta$ is the weight parameter that adjusts the relative importance of SNR and INR, which is chosen around $1$. Large $\zeta$ (e.g. $\zeta>1$) indicates more focus on mitigating INR rather than enhancing SNR, which leads to more conservative transmission power. This benefits the UEs with poor channel states due to the reduced interference, while constraining the transmission power of UEs with good channel states that prevents them from achieving better throughput. It works the other way around when we choose small $\zeta$ (e.g. $\zeta <1$). Tending to suppress strong interference, we initially select $\zeta$ to be $1.3$. As shown in Fig. \ref{Sample_Utility_Sum}, the weighted sum throughput is maximized at a unique transmission power, i.e., the balance we choose between SNR and INR. 

\IncMargin{1em}
\begin{algorithm}[h]
\SetKwFunction{CandidateUser}{CandidateUser}
\SetKwData{NULL}{NULL}
\SetKwInOut{Input}{input}
\SetKwInOut{Output}{output}

 \textbf{Given} $l= -10$ dBm, $r = P_{\max}$, \text{tolerance} $\epsilon=0.1$\;
\uIf{$r-l < \epsilon$} {
$m:= P_{\max}$;
}
\Else{ 
 \While{$r-l \geq \epsilon$}{
   
       $m:= (l+r)/2$\;
       \uIf{$R_S'(m) + \zeta \cdot R_I'(m)< 0$}{
       $l:=m$; 
       }
       \Else{ 
      $r:=m$;
      }
    }
}

\Return{$P:=m$.} 
 
\caption{Bisection method for solving (\ref{Gradient_0}).}\label{alg_Bisection}
\end{algorithm}
\vspace{-0.2in}
\DecMargin{1em}

It is easy to show that Problem (\ref{U_Maximize}) is a one-dimensional quasi-convex optimization problem, and thereby can be obtained by solving the following problem:
\begin{eqnarray}
R_S'(P) + \zeta \cdot R_I'(P)= 0. \label{Gradient_0}
\end{eqnarray}
We use bisection method to find the solution of (\ref{Gradient_0}), with the detailed steps presented in Algorithm \ref{alg_Bisection}. This algorithm is easy to implement in software, and the number
of iterations for convergence is no more than 10. We note that Algorithm 1 is a fully distributed algorithm, where each UE can choose its transmission power independently.

\section{Performance Evaluation on System-level Simulation Platform}\label{Simulation}

\subsection{Simulation Platform Configuration}
We consider an LTE uplink cellular network \cite{36.814}, where the BSs are located on
a typical hexagonal lattice including $19$ BS sites, $3$ sectors per site. Each sector is regarded as a cell. The minimal distance between two neighboring sites is 500 meters (Macrocell) \cite{36.814}. The UEs are uniformly distributed in the entire network area. The distance from a UE to a nearest BS is no smaller than 35 meters \cite{3GPP25.996}. Each UE has a single antenna, and the BSs are equipped with 2 antennas per sector. The wireless channel coefficients and BS antenna pattern are generated by following the SCM model for Urban Macro environments in 3GPP TR 25.996 \cite{3GPP25.996}, where 3D antenna pattern and Rayleigh fading \cite{Zheng03} are adopted. We set the maximum doppler shift frequency at $7$ Hz according to moving speed $3$ km/h and carrier frequency $2.5$ GHz. The receivers employ non-CoMP maximum mean square error (MMSE) estimation \cite{Tse05} with interference rejection combination (IRC) techniques \cite{Leost12}. To obtain the channel coefficients, we estimate the pilots, i.e., demodulation reference signals (DMRS) \cite{36.213}, with DFT-based estimation \cite{Huang10}. 
Adaptive Transmission Bandwidth (ATB) \cite{ATB} non-CoMP packet scheduling scheme is implemented. The frequency bandwidth is 10MHz, and the noise figure of each BS receiver is 5 dB \cite{36.213}.
We set a uniform penetration loss of $20$ dB \cite{3GPP25.996} for all users. 
The delay of control signaling is uniformly set as $6$ ms (i.e., 6 time slots) \cite{36.321}. The wrap around technique is employed to avoid the border effect. The parameters of our system level simulation platform are listed in Table \ref{tab2}.

We use the proportional-fair (PF) policy \cite{Tse05} in stochastic network control in which the weight of UE $u$ used in the ATB scheduler at time-slot $t$ is
\begin{eqnarray}
\frac{(r_u[t])^\alpha}{(\bar{r}_u)^\beta},  \label{PF_eq}
\end{eqnarray}
where $r_u[t]$ and $\bar{r}_u$ denote the UE $u$'s potentially achieved data rate in time slot $t$ and long-term average data rate, respectively. We set the two associated parameters at: $\alpha =1$, $\beta = 1$. 

We compare C\&B with three reference policies: One policy is the widely used fractional power control (FPC) scheme, which determines the transmission power $P^{FPC}$ of UE $u$ for each resource block by \cite{36.213}:
\begin{eqnarray}
P^{FPC} = \min(P_{\max}, P_0^{FPC} + \kappa \cdot \text{PL}),   \label{FPC_eq}
\end{eqnarray}
where $P_{\max}$ is the maximum power constraint, $P_0^{FPC}$ is the default transmission power, and $\text{PL}$ is the large-scale path loss from UE $u$ to its serving cell. The values of the parameters are $P_{\max} = 23$ dBm, $P_0^{FPC} = -87$ dBm and $\kappa = 0.8$. The second policy is Max Power, which sets all UEs at their maximum allowable transmission power $P_{\max}$. The third reference policy is the coordinated Reverse Link Power Control (RLPC) scheme, where the transmission power $P^{RL}$ is decided by \cite{Rao07}:
\begin{eqnarray}
P^{RL} = \min\big(P_{\max}, P_0^{RL} + \phi \cdot \text{PL} + (1-\phi) \cdot \text{PL}_{\min} \big),   \label{RLPC_eq}
\end{eqnarray}
where $\text{PL}_{\min}$ denotes the measured minimum path loss from $u$ to its neighboring cells. The rest parameters are selected as $P_0^{RL} = -102$ dBm and $\phi = 0.8$. The parameters of C\&B are chosen as we discussed in Section \ref{Main Design}.

\begin{table}

\centering
\caption{SYSTEM LEVEL SIMULATION PARAMETERS}\label{tab2}
\resizebox{\columnwidth}{!}{
\begin{tabular}{l|l}
 \hline
Parameter& Setting\\ 
\hline
Deployment Scenario & 19 BS sites, 3 sectors (cells) per site, \\
&wrap-around \\
Inter-site Distance & 500m (Macrocell)\\
System Bandwidth & 10MHz [50 PRBs, 2 used for control] \\
Avg. UEs per Cell & 10\\
UE/BS Antennas & 1/2 per cell\\
Distance-dependent Path Loss& According to 3GPP 36.814 \cite{36.814}\\
Shadowing Standard Deviation & 8 dB \\
Antenna Pattern & 3D\\
Penetration Loss & 20 dB\\
Scheduling Decision Delay & $6$ ms\\
Target BLER & $10\%$\\
Traffic Model & Fully backlogged queues \\
Scheduling Algorithm & ATB\\
Power Control (PC) scheme & FPC, Max Power, RLPC, C\&B \\
Stochastic Network Control Scheme & Proportional Fair (PF)\\
Link Adaptation & AMC, based on 3GPP TS 36.213 \cite{36.213}\\
BS Receiver Type & IRC MMSE\\
Channel Estimation & DFT-based Estimation\\
Maximum Doppler Shift & 7Hz \\
$\alpha$ (PF) & 1\\
$\beta$ (PF) & 1\\
$P_{\max}$ & $23$ dBm\\
$P_0^{FPC}$ (FPC)& $-87$ dBm\\
$\kappa$ (FPC)& 0.8 \\
$P_0^{RL}$ (RLPC) & $-102$ dBm\\
$\phi$ (RLPC) & 0.8\\
 \hline
\end{tabular}
}
\vspace{-0.2in}
\end{table}

\subsection{Simulation Results}
We compare the performance of different power control schemes in terms of three key metrics: cell average throughput (i.e. sum average throughput per cell), cell edge throughput and power efficiency. In particular, cell-edge throughput is defined as the $5^{th}$ percentile throughput performance among all UEs, denoted as $5\%$-Edge, which is widely used in evaluating the performance of UE fairness \cite{Furuskar08,Castellanos08,Coupechoux11}.
\noindent
\begin{figure*}[htp!]
\vspace{-0.1in}
\centerline{
\subfloat[Time-average received SNR.]
{\includegraphics[width=2in]{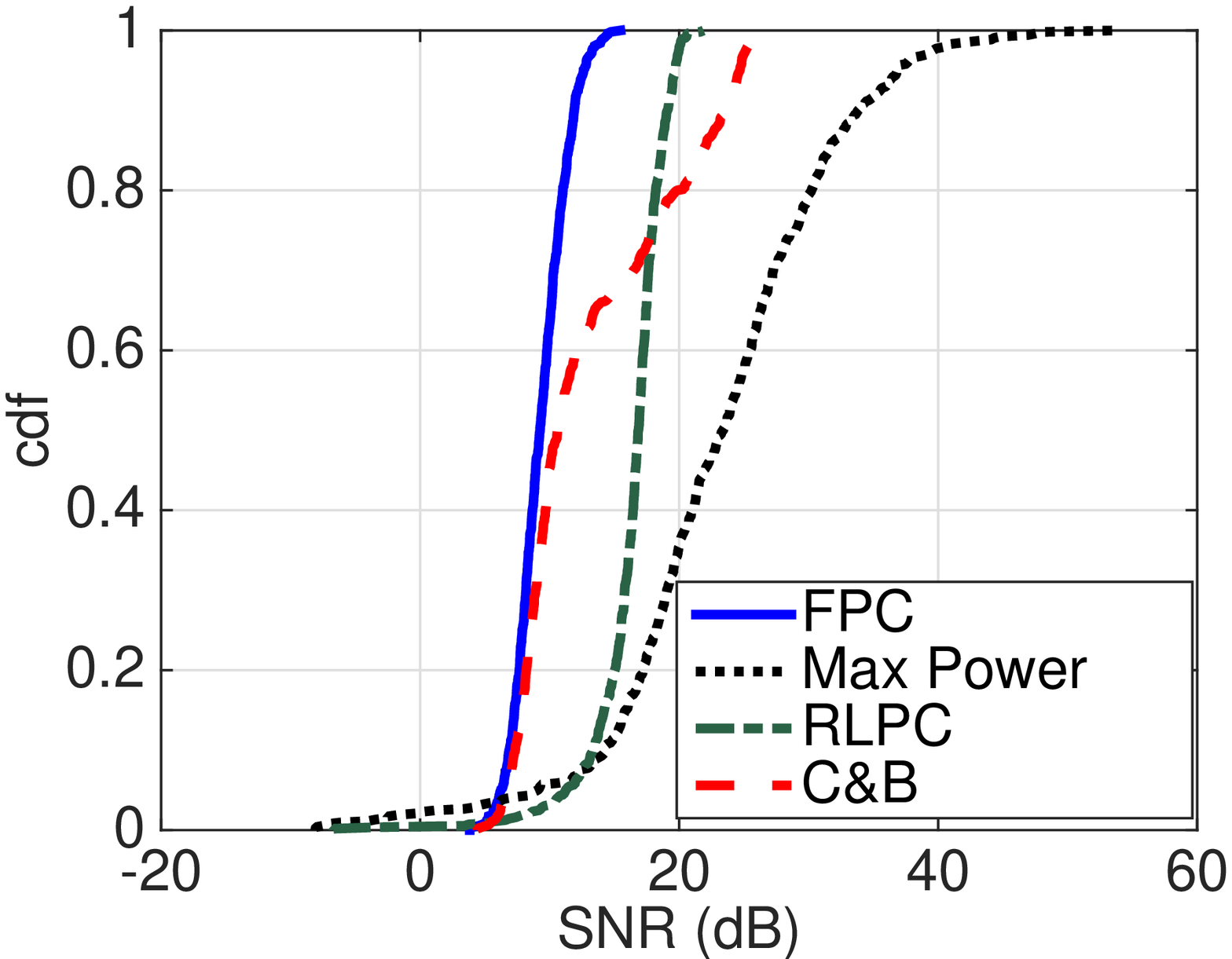}
\label{SNR_500m_7Hz}}
\subfloat[Time-average experienced IoT.]
{\includegraphics[width=2in]{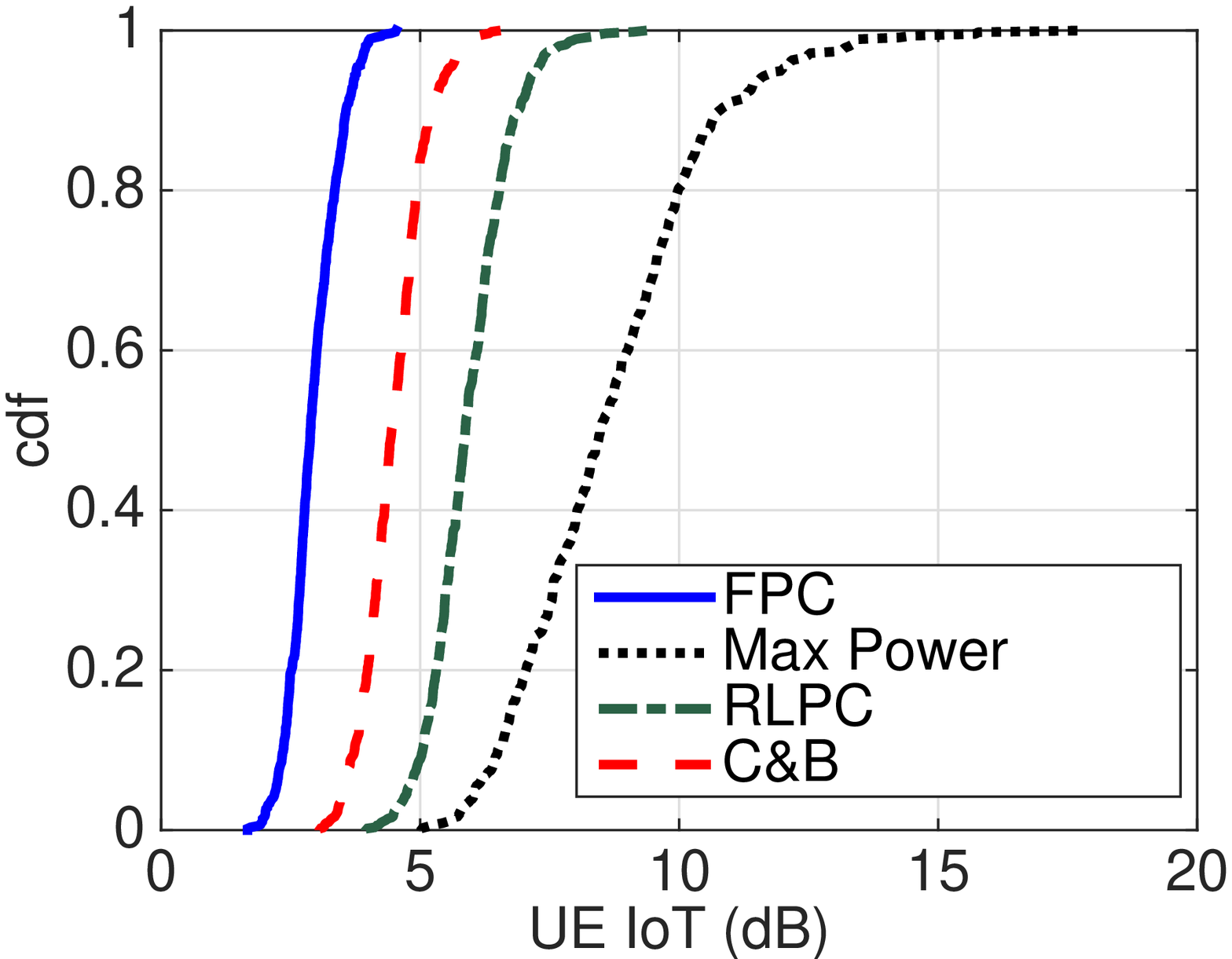}
\label{IoT_500m_7Hz}}
\subfloat[Time-average throughput. ]
{\includegraphics[width=2in]{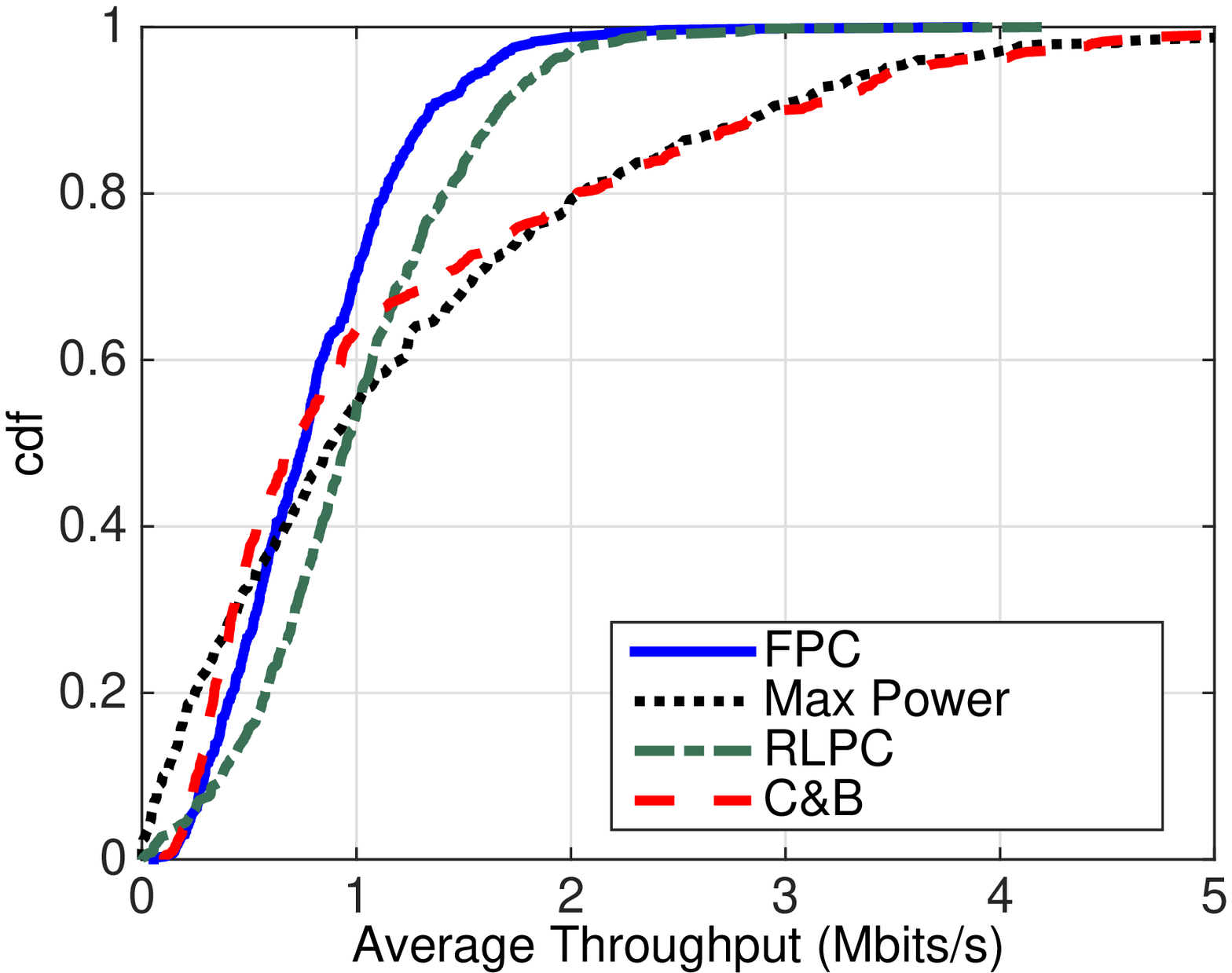}
\label{Throughput_500m_7Hz}}
}
\caption{Simulation results in Macrocell scenario.}
\vspace{-0.2in}
\end{figure*}

\subsubsection{Performance Comparison in Macrocell Scenario}
First, we investigate the performance in enhancing received signal strength, which is evaluated by time-average SNR as illustrated in Fig. \ref{SNR_500m_7Hz}. It can be seen that C\&B, with the weight parameter set as $\zeta = 1.3$, is able to significantly boost UEs' time-average SNR over FPC. Particularly, more than $40\%$ UEs have gained at least $3.5$ dB and approximately $20\%$ UEs  even achieve more than $10$ dB SNR increase. C\&B also improves the SNR of the top $20\%$ UEs compared to RLPC, which greatly benefit the UEs with good channel conditions. It is clear that Max Power has the best SNR performance due to the maximum transmission power. Nevertheless, it simultaneously incurs severe interference which is quite undesirable.

Next, we switch our attention on the interference mitigation performance, which is represented by the time-average IoT as shown in Fig. \ref{IoT_500m_7Hz}. Compared with FPC, C\&B only slightly increase the UEs' average IoT by $1.5$ dB. On the other hand, C\&B performs a lot better than Max Power and shows clear advantage over RLPC in suppressing the inter-cell interference.

Now we concentrate on the throughput performance, which is the comprehensive result of signal enhancement and interference mitigation. In comparison with FPC and RLPC,  C\&B noticeably improves the throughput of UEs with good channel condition (the top $30\%$), as illustrated in Fig. \ref{Throughput_500m_7Hz}. In particular, the top $20\%$ UEs have even achieved throughput gain of at least $1$ Mbits/s. As for UEs with poor channel condition (bottom $10\%$), C\&B shows great advantage over Max Power.

Finally, we present C\&B's advantages over all other candidate power control schemes based on the detailed performance summary in Table \ref{Table_PerformanceSummary}. C\&B shows desirable advantage over FPC by providing $51.9\%$ improvement on cell average throughput, while keeping the same $5\%$-Edge throughput performance and only dropping the power efficiency by $48\%$. Further, C\&B beats Max Power in significantly improving $5\%$-Edge throughput and power efficiency by $156\%$ and $5,716\%$ respectively,  with slightly increased cell average throughput. Comparison with a CoMP power control scheme called RLPC, C\&B achieves appreciable gains in cell average throughput and power efficiency respectively by $25.1\%$ and $71.2\%$, and simultaneous keeps $5\%$-Edge throughput slightly improved. 

Note that the performance improvement of C\&B is solely obtained by power control. One can further incorporate coordinated transceiver and scheduling techniques to achieve even higher gain. For example, it is known that CoMP receiving techniques can enhance the edge throughput significantly \cite{Suh11}. As shown in Table \ref{Table_PerformanceSummary}, the throughput gain of C\&B is more evident in the high throughput UEs. Hence, additional improvement is promising by combining C\&B with other CoMP techniques.

\subsubsection{Tradeoff between cell average throughput and cell edge throughput}

Recall the utility maximization problem (\ref{U_Maximize}), we can vary the weight parameter $\zeta$ to adjust the balance between received signal enhancement and interference mitigation. As we gradually decrease $\zeta$ from $1.3$ to $0.7$, the cell average throughput keeps improving in pair with the continuously degrading $5\%$-Edge throughput, as summarized in Table \ref{Table_DifferentWeight}. This tradeoff between average and edge throughput can be explained as follows: (a) the UEs with good channel conditions contribute most part of the average throughput. Increasing such UEs' transmission power by decreasing $\zeta$ greatly helps them to achieve stronger received signal, which results in their throughput gains. Such gains lead the cell average throughput to improve. (b) In contrast, boosting transmission power incurs stronger inter-cell interference, which especially jeopardizes those vulnerable UEs with poor channel states. As a result, the $5\%$-Edge throughput is decreased. 

\begin{table}[h!]
\centering
\caption{Throughput performance comparison between different weight factor $\zeta$ selections.}
\resizebox{\columnwidth}{!}{
\begin{tabular}{|| c || c |  c | c | c ||}
 \hline
       & $\zeta = 1.3$ & $\zeta = 1.1$ & $\zeta = 0.9$ & $\zeta = 0.7$\\ 
 \hline
Average Throughput (Mbits/s) & $12.23$ & $12.41$ & $12.95$   & $13.17$   \\
\hline
$5\%$-Edge Throughput (Mbits/s) & $0.23$ & $0.21$ &   $0.18$  & $0.15$ \\ 
\hline
\end{tabular}
}
\label{Table_DifferentWeight}
\end{table}

Note that when $\zeta = 0.7$, C\&B is remarkably better than the Max Power policy in terms of both average and edge-throughput. In addition, when $\zeta = 1.3$, C\&B is significantly better than the FPC and RLPC in term of cell average throughput. Therefore, one can adjust the weight parameter in order to adapt different system requirements.

\section{Conclusion and Future Work}\label{Conclusion}
We investigate how to use limited large scale CSI to achieve the throughput gain of CoMP through the design of a power controller named C\&B. The optimal power control by itself is an NP hard problem, which has been open up to date. Further, as very limited coordination is possible in the open-loop mode, the power controllers of different BSs are not allowed to communicate. C\&B satisfies all practical constraints in cellular systems, and can significantly improve the average and edge throughput over existing power control schemes with very low complexity and almost no cost. Further throughput enhancement is promising by combining C\&B with other coordinated transceiver and scheduling techniques.



\end{document}